\documentclass[a4paper]{spie}  

\usepackage{amsmath,amsfonts,amssymb}
\usepackage{graphicx}\graphicspath{{figures}}
\usepackage[colorlinks=true, allcolors=blue]{hyperref}
\usepackage{bm}
\usepackage{esint}

\title{A numerical integration scheme for vectorised phase-space of one-dimensional collision-free, electrostatic systems}

\author[a]{Allen Lobo}
\author[a]{Vinod Kumar Sayal}
\affil[a]{Department of Physics, Sikkim Manipal Institute of Technology, Sikkim Manipal University, Sikkim-737136, India.}

\authorinfo{Further author information: (Send correspondence to A.L.)\\A.L.: E-mail: allen.e.lobo@outlook.com, Telephone: +91 9031 953805}

\begin{document} 
\maketitle

\begin{abstract}
The kinetic analyses of many-particle soft matter often employ many simulation studies of various 
physical phenomena which supplement the experimental limitations or compliment the theoretical 
findings of the study. Such simulations are generally conducted by the numerical integration techniques of the governing equations. In the typical case of collisionless electrostatic systems such as electrostatic plasmas, the Vlasov-Poisson (VP) equation system governs the dynamical evolution of the particle phase-space. The one-dimensional position-velocity (1D-1V) particle phase-space, on the other hand, is known to exhibit direct analogy with ordinary two-dimensional fluids, wherein the Vlasov equation resembles the fluid continuity equation of an in-compressible fluid. On the basis of this fluid-analogy, we present, in this work, a new numerical integration scheme which treats the 1D-1V phase-space as a two-dimensional fluid vector space. We then perform and present analyses of numerical accuracy of this scheme and compare its speed and accuracy with the well-known finite splitting scheme, which is a standardised technique for the numerical Vlasov-Poisson integration. Finally, we show some simulation results of the 1D collisionless electrostatic plasma which highlight the higher speed and accuracy of the new scheme. This work presents a fast and sufficiently accurate numerical integration technique of the VP system which can be directly employed in various simulation studies of many particle systems, including plasmas.
\end{abstract}

\keywords{Computational Plasma Physics, Numerical Techniques, Numerical Vlasov Integration, ANI scheme}

\section{INTRODUCTION}\label{sec:intro}  
The kinetic theory of matter is well-known to be a meticulous approach to understand various aspects of the dynamics of many-particle systems. This approach employs a statistical distribution of the mechanical states of various particles in the system in order to explain the macroscopic phenomena which occur in it. Plasmas, which are high-energy, partially or completely ionised states, are not unknown to the kinetic approach -- their dynamics have been assiduously explored using the plasma kinetic theories\cite{VanKampen1955, Landau1946OnPlasma, Manfredi1997, Brodin1997NonlinearDamping, Sayal1990ElectronPlasma} which includes certain phenomena that can not be addressed by the magneto-hydrodynamics approach. In the plasma kinetic theory, the dynamical evolution of the system is exhibited by the well-known Boltzmann equation. In the hot plasma case, collisions and inter-particle interactions can be ignored and the equation modifies into the Vlasov equation\cite{Vlasov1938},
\begin{equation}\label{vlassov_eqn}
    \frac{\partial f}{\partial t} + \vec{v}\cdot\frac{\partial f}{\partial \vec{r}
} + \vec{a}\cdot\frac{\partial f}{\partial \vec{v}}=0.\end{equation}
In the above equation, $f(\vec{r},\vec{v},t)$ represents the probability distribution function of the plasma species in its position-velocity phase-space. The terms $\vec{r}$, $\vec{v}$ and $\vec{a}$ represent position, velocity and acceleration vectors and $t$ represents time, respectively. The Vlasov equation (\ref{vlassov_eqn}) describes the dynamical evolution of the system in terms of its phase-space distribution function. This evolution is regulated by the presence of field interactions in the system, which collectively contribute to the acceleration term $\vec{a}$. In the case of purely electrostatic interactions, this acceleration term can be represented by the contribution of the electric field $\vec{E}$, using the Poisson equation,
\begin{equation}\label{Poisson_eqn}
   \vec{a} = \frac{q}{m}\vec{E}, \quad \vec\nabla_{\vec{r}}\cdot \vec{E}=-\nabla_{\vec{r}}^2 \phi = \frac{1}{\varepsilon_0} \sum_{i=1}^N q_i\int f_i(\vec{r},\vec{v})dv. 
\end{equation}
In the above equation, $\varepsilon_0$ is the spatial electrostatic permittivity, $\phi$ is the electrostatic potential, $q_i$ is the charge of the plasma species and $m$ is the mass of the species. Equations (\ref{vlassov_eqn}) and (\ref{Poisson_eqn}) collectively describe the Vlasov-Poisson (VP) system, which governs the dynamical evolution of the phase-space kinetics of the plasma species. The kinetic approach encompasses various linear and nonlinear phenomena which occur in plasmas, including wave propagation, instabilities and structure formations. The Vlasov equation (\ref{vlassov_eqn}) exhibits remarkable similarity, in its form, to the continuity equation of an ordinary in-compressible fluid. This similarity has been explored by some authors \cite{Berk1970PhaseObservations, schamel1986electron, schamel2012cnoidal} and it was recently shown\cite{Lobo2023} that this analogy extends beyond the Vlasov equation, that is, a 1D-1V phase-space and a two-dimensional conventional fluid surface exhibit similarities in terms of the evolution dynamics, flow-like behaviour and deformations due to external fields.

Kinetic analyses of the plasma phenomena are often accompanied by numerical simulations of the system\cite{Forslund1985}, which, apart from visualising and showcasing the studied phenomena, also exhibit the dynamics of the field waves which may not be observable in experimental studies, and also help in validating the study and its findings. The simulation is generally conducted by numerically integrating the VP equation system. Various numerical schemes have been developed and analysed in this regard \cite{Watanabe2001, Aminmansoor2015HybridModel, turikov1978computer, Filbet2001, morse1969one, Colombi2008Vlasov-Poisson:Revisited, Filbet2003}. One of the well-known benchmark numerical schemes, developed by Cheng and Knorr (1976)\cite{Cheng1976}, known popularly as the finite splitting (FS) scheme, has been used in various numerical investigations of nonlinear processes in plasma physics\cite{Filbet2001, Filbet2003}. The FS scheme employs a leap-frog integration scheme of the Vlasov-equation along position and velocity spaces, each at different, half-time steps. These steps, each marked by asterisks $(^*)$, are as follows:
\begin{equation}\label{cheng_knorr_scheme}
    f(x,v_x,t)\rightarrow f^*(x-v_x\Delta t/2, t)
    \rightarrow f^{**}(x,v_x-\Delta t q\vec{E}/m,t)\rightarrow  f^{***}(x-v_x\Delta t/2, t) = f(x,v_x,t+\Delta t).
\end{equation}
In the above numerical equation, $\Delta t$ represents the time-step value and $m$ represents the mass of the plasma species. This 3-step integration scheme of the Vlasov equation employs various interpolation schemes for shifting the distribution function along respective position and velocity axes. In the above equation (\ref{cheng_knorr_scheme}), each arrow $(\rightarrow)$ indicates this interpolating shift of the phase-space distribution function (df). This means that for the evolution of the phase-space by one time-step, a three-step algorithm should be implemented, which may prove to be an overburden on computation systems, when dealing with huge systems. This is due to the dependency of the accuracy of this scheme on the grid-spacing\cite{Courant1928}. Therefore, large-sized systems with high sensitivity to numerical accuracy might result in longer simulation times and slower data generation. This limitation serves as the chief motivation of this work.

In this article, we present a new numerical algorithm based on vectorisation of the plasma phase-space. By vectorisation, it is connoted that the phase-space df is observed as a three-dimensional vector field, with independent evolving components along each axis of the dynamical phase-space: position $(x)$, velocity $(v_x)$ and time $(t)$. We then expand the evolution of the phase-space df separately along each axis and present an agile, numerical integration scheme which evolves the phase-space by one-time step in one numerical step, which significantly reduces the computational time of the evolution equation when compared to the FS scheme. We then study and present the stability of our code and compare its speed and accuracy with the FS scheme for some well-known plasma phenomena. This article is planned as follows: in section \ref{sec2}, the three-dimensional outlook of the time-evolving dynamical phase-space is discussed and the phase-space is described as a three-dimensional vector space. Phase-space evolution is then exhibited as a flow of the phase-space hydrodynamic behavior along the three dimensions independently. In section \ref{sec3}, using this idea of independent flow-like evolution of the phase-space, a new numerical scheme is developed, and its stability and accuracy is discussed. In section \ref{sec4}, the new numerical scheme is employed to simulate various plasma phenomena, and its numerical accuracy and speed are tested against the FS scheme. Finally, section \ref{sec5} is devoted to the concluding remarks. This article presents a faster and accurate numerical scheme which can be used in computational set-ups to observe and study various linear and nonlinear plasma phenomena with much less computational burden than the existing techniques.

\section{Evolving phase-space as a three-dimensional vector field}\label{sec2}
The 1D-1V particle phase-space has often been compared\cite{schamel2012cnoidal, Berk1970PhaseObservations} to a two-dimensional fluid surface, by observing the well-known analogy between the Vlasov equation and the continuity equation. While the latter represents the nature of flow and conservation of fluid density, the former states that a particle in its phase-space moves only along a trajectory where-in its probability density remains a constant, that is, the total time derivative of the phase-space df is zero, specifically in the collisionless case. This analogy, however, lies beyond the Vlasov-Continuity equations, as shown by Lobo and Sayal\cite{Lobo2023}. It is shown that the particle phase-space also possesses its own vorticity field, and experiences a statistical pressure due to non-uniform probability distributions, and can also form vortices\cite{Guio2003, hutch2017, Lobo2023} and other coherent structures. Kinetic hydrodynamic theories can then be developed in order to study these fluid-analogous structures in the phase-space\cite{Lobo2023}.

\begin{figure}[!ht]
    \centering
    \includegraphics[width=0.7\linewidth]{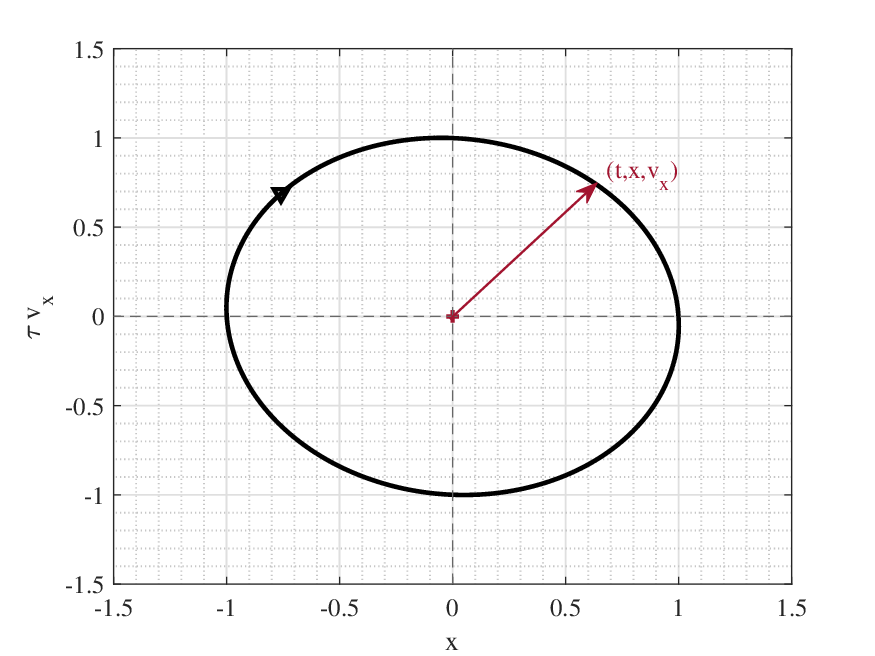}
    \caption{Phase-space trajectories of a simple harmonic oscillator. A two-dimensional $(x,\tau v_x)$ phase-space portrait representing motion of the particle along the phase-space trajectory. At any given time, the state of the particle lies in the phase-space trajectory (indicated by the red arrow). However, the exact position of the particle in its phase-space can not be found at a given time in this representation.}
    \label{fig:2doscillation}
\end{figure}

\begin{figure}[!ht]
    \centering
   \includegraphics[width=0.7\linewidth]{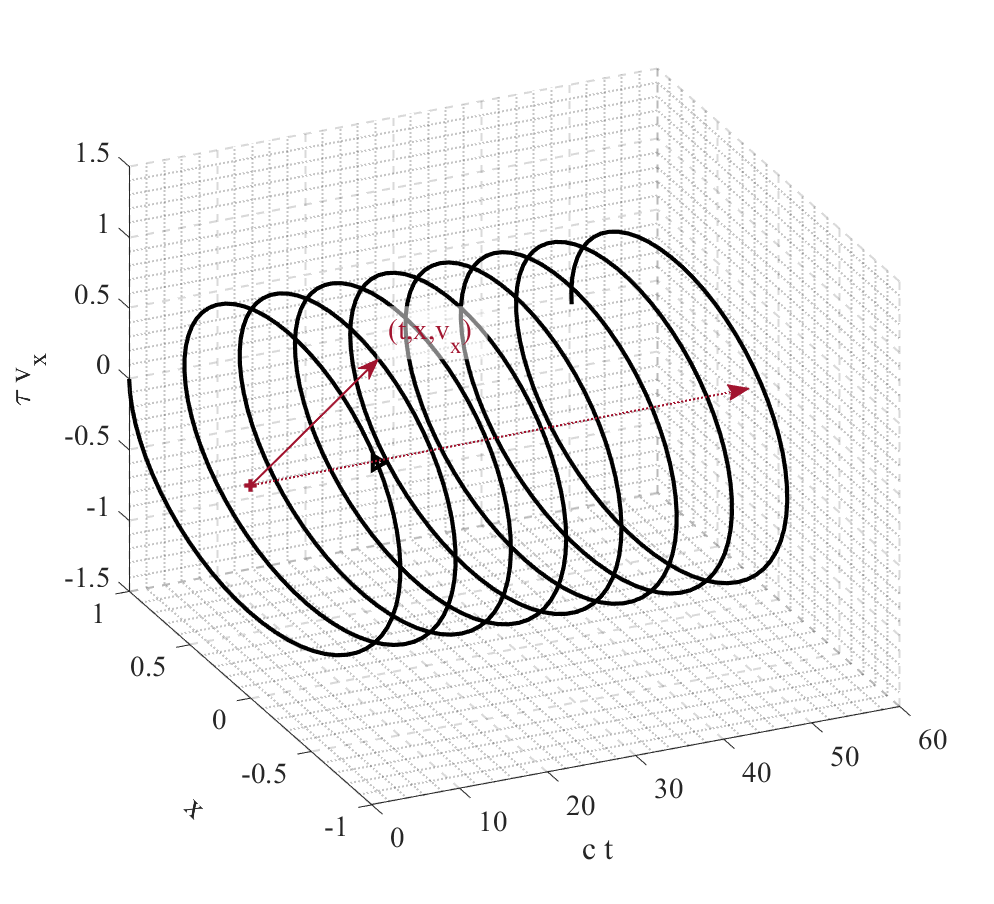}
    \caption{Phase-space trajectories of a simple harmonic oscillator, in a three-dimensional representation $(ct,x,\tau v_x)$. The mechanical state of the particle at a specific time can be easily located in this representation (red, continuous arrow). The red, dotted arrow represents the unidirectional flow of time, from the origin $(ct=0,x=0,\tau v_x=0)$. }
    \label{fig:3doscillation}
\end{figure}

The dynamical evolution of the phase-space is described by the time-dependence of the phase-space df $f(x,v_x,t)$. While this time-dependence can be both extrinsic or intrinsically through the space and velocity coordinates, it is well-known that the evolution of the phase-space df can be described by the motion of the particle along a phase-space trajectory which conserves this density, that is,
\begin{equation}
    \frac{df(x,v_x,t)}{dt}=0, \quad\text{for collisionless case.}
\end{equation}
At a certain time, an infinitesimal phase-space volume element $(\delta V=\tau \delta x \delta v_x)$ can be assigned a phase-space coordinate set $(x,\tau v_x)$, where $\tau$ is a characteristic time of the system. For plasmas, it is the inverse of the plasma frequency $\omega_p = \sqrt{n_0q^2/\varepsilon_0 m}$, where $n_0$ is the unperturbed spatial particle density. Therefore, the spatio-temporal location of the phase-space volume element becomes $(ct,x,\tau v_x)$, where $c$ is a characteristic speed, at which information travels in time, and is closely related to the Courant-Friedrichs-Lewy (CFL) number \cite{Courant1928}. Usually, $c$ is is taken to be equal to $1$. The temporal-phase-space evolution is then observable in a three-dimensional phase-space, representing a one-dimensional system. As an example of the same, we present the case of a simple harmonic oscillator with unit angular frequency --
\begin{equation}
    \frac{dv_x}{dt}= -x,\qquad \frac{dx}{dt}=v
_x. 
\end{equation}
Fig, \ref{fig:2doscillation} shows the phase-space trajectory of the harmonic oscillator, for a total of eight periods. At any instant, the mechanical state of the particle is represented by the coordinate set $(x,\tau v_x)$. 
Fig. \ref{fig:3doscillation} showcases the same phase-space, evolving with time in three-dimensions. While in both cases, the information relayed is the same, the three-dimensional representation of the phase-space is a better representation of the dynamically evolving system.

Having described the phase-space in a three-dimensional set-up of position, velocity and time, the state of a phase-pace volume element $(\delta V)$ can be described in a three coordinate set $(ct, x, \tau v_x)$. Therefore, at any instant, the position vector $\bm{\Gamma}$ of the $\delta V$ element can be described as:
\begin{gather}
    \bm{\Gamma} = ct\hat t + x\hat x + \tau v_x \hat{v}_x.\label{Gamma}
    \intertext{Differentiating w.r.t. time, we get the velocity field $\bm{u}$ of the volume element $\delta V$,}
    \bm{u} = c\hat t + v_x \hat x + \frac{q}{m}\vec{E} \hat{v}_x.\label{u}
\end{gather}
The phase-space velocity field $\bm{u}$ defined in equation (\ref{u}) describes the evolution of the particle phase-space in terms of a flow-like behaviour of the phase-space df in phase-space and time. This hydrodynamic behaviour of the phase-space in a time-independent case was introduced earlier by Berk, Nielson and Roberts\cite{Berk1970PhaseObservations} and later by Lobo and Sayal\cite{Lobo2023}. While the velocity components along $\hat x$ and $\hat{v}_x$ directions are known from the system and system-specific interacting fields\cite{Lobo2023}, the phase-space velocity component along the time axis $(\hat t)$ describes the rate of evolution of the dynamical system, as follows:

For the system evolves for a time $\Delta t$, the phase-space must also evolve accordingly,
\begin{gather}\allowdisplaybreaks
    t \xrightarrow{t+\Delta t} t+\Delta t, \label{delta_t}\\
   x \xrightarrow{t+\Delta t} x + \hat{x}\cdot\bm{u}\Delta t = x + \Delta x\bigr|_{t}^{t+\Delta t},\label{delta_x}\\
   \tau v_x \xrightarrow{t+\Delta t} \tau v_x + \tau\hat{v}_x\cdot\bm{u}\Delta t = \tau v_x + \tau \Delta v_x\bigr|_{t}^{t+\Delta t},\label{delta_v}\\
  \therefore  f(ct,x,\tau v_x)\xrightarrow{t+\Delta t} f(ct+c\Delta t, x + \Delta x\bigr|_{t}^{t+\Delta t}, \tau v_x + \tau \Delta v_x\bigr|_{t}^{t+\Delta t}).
    \intertext{Therefore, $c$ is the rate at which the phase-space fluid evolves along the time axis $\hat{t}$. If $c$ is less than $1$, it would imply that the phase-space evolves slower than the information propagating along-with the evolving time. This means that there would exist an asymmetry in the development of the phase-space df itself and its coordinates $x$ and $\tau v_x$ which develop along-with time. On the other hand, if $c$ is greater than $1$, it implies that the speed at which the information associated with the system's evolution travels faster than the system itself. In a physical system, due to the idea of causality, this condition becomes meaningless. However, this condition may be allowed in a numerically simulated system following an implicit time-integration scheme. Therefore,}
    c\geq 1.\label{cvalue}
\end{gather}
The restriction imposed on the value of $c$ in equation (\ref{cvalue}) is also in agreement to the CFL condition\cite{Courant1928}, and $c$ itself is therefore identical to the Courant number.

Having described the dynamically evolving phase-space as a three-dimensional vector space, the Vlasov equation can be rewritten in a vector form as the divergence of the phase-space flux density $\bm{\mathcal{J}}$, which can be defined as the phase-space velocity field $\bm{u}$ times the phase-space df $f(ct,x,\tau v_x)$.
\begin{gather}
    \bm{\nabla} = \frac{1}{c}\frac{\partial}{\partial t}\hat{t} + \frac{\partial}{\partial x}\hat x + \frac{1}{\tau}\frac{\partial}{\partial v_x}\hat{v}_x.\\
    \bm{\mathcal{J}} = f\bm{u} = cf\hat{t} + v_xf\hat{x} - \frac{q\tau}{m}\frac{\partial \phi}{\partial x}f\hat{v}_x,\label{j_Value}
    \intertext{where, the electric field $\vec{E}$ is substituted by the electric potential gradient $(-\partial \phi/\partial x)$. The Vlasov equation then takes the form,}
    \bm{\nabla}\cdot\bm{\mathcal{J}} = 0.\label{vlasoveqninvectorform}
\end{gather}
 Equation (\ref{vlasoveqninvectorform}) presents the Vlasov equation in a non-divergent, differential form. It is well-known from conventional vector calculus that equation (\ref{vlasoveqninvectorform}) can also be represented in an integral form as follows:
 \begin{equation}\label{integral_vlasoveqn}
     \oiint \bm{\mathcal{J}}\cdot d\bm{S} = 0.
 \end{equation}
In the above equation, $d\bm{S}$ represents surface elements in the three dimensional dynamical phase-space, and is equal to
\begin{equation}
    d\bm{S} =\tau dxdv_x\hat{t}+ c\tau dt dv_x\hat{x}+ cdtdx\hat{v}_x .
\end{equation}
The integral form of the Vlasov equation (\ref{integral_vlasoveqn}) presents the well-known\cite{Lobo2023} solenoidal nature of the phase-space flux density field. It states that as the system evolves in time, there is no net flux of the phase-space fluid across a closed dynamical phase-space volume $\int_{\delta t}\oiint \delta V\cdot  c\delta t$. 

\begin{figure}[!ht]
    \centering
    \includegraphics[width=0.7\linewidth]{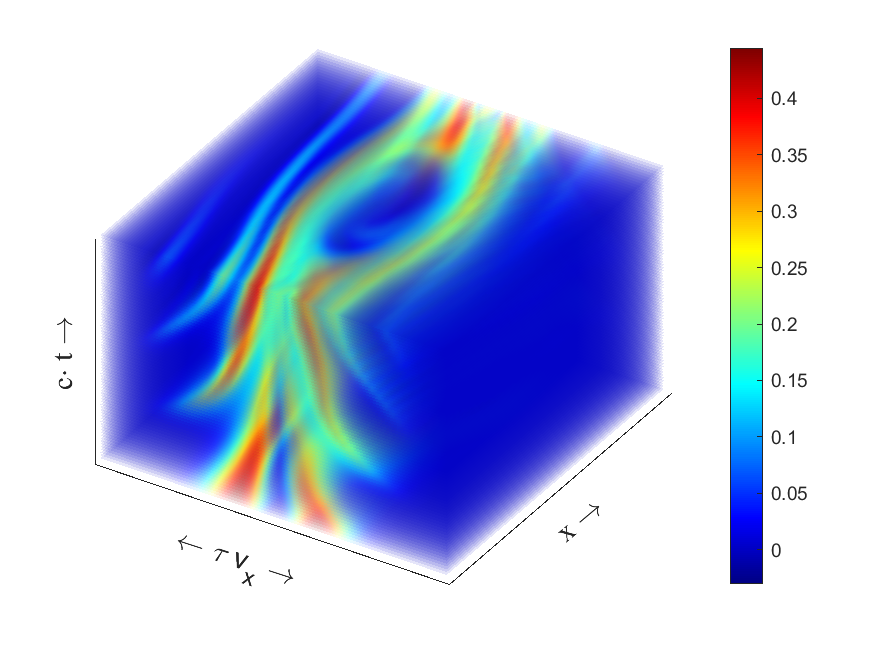}
    \caption{Evolution of the plasma dynamic phase-space. Phase-space df portrait shown during the evolution of an initial two-stream plasma into a phase-space vortex. Color gradient representing magnitude of the phase-space density $f(ct,x,\tau v_x$ and its evolution with time.}
    \label{fig:3d_hole_f}
\end{figure}

\begin{figure}[!ht]
    \centering
    \includegraphics[width=0.7\linewidth]{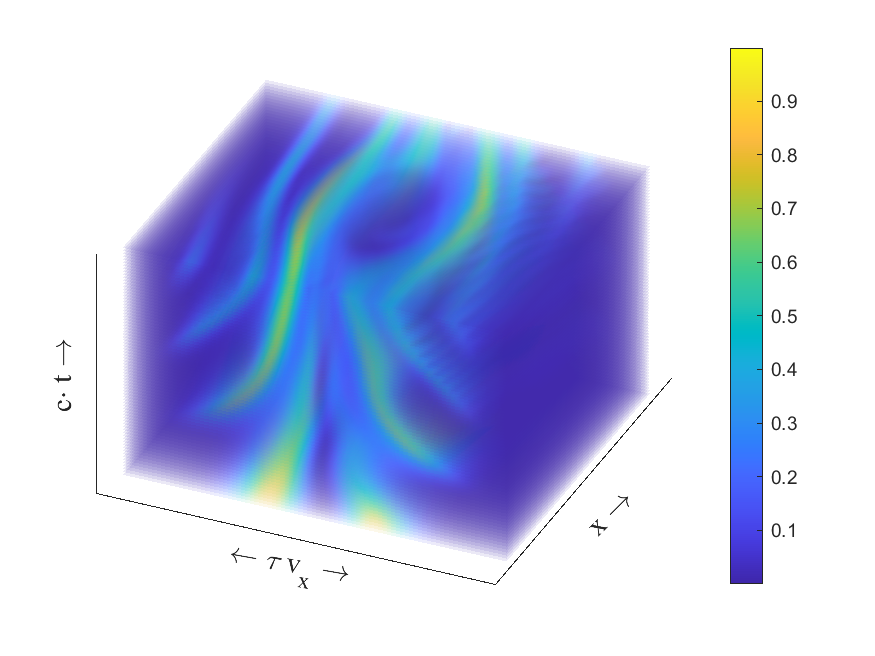}
    \caption{Evolution of the plasma dynamic phase-space flux density magnitude $|\bm{\mathcal{J}}|$ from an initial two-stream plasma into a phase-space vortex, with magnitudes represented bu the color gradient.}
    \label{fig:3dhole_j}
\end{figure}

Therefore. the dynamical evolution of the particle phase-space can be conveniently observed as a kinetic-hydrodynamic flow along three independent coordinates, of time, length and speed, in a three-dimensional vector space. Figs. \ref{fig:3d_hole_f} and \ref{fig:3dhole_j} portray the temporal evolution of the dynamical phase-space in terms of the phase-space df and the magnitude of the flux density field $\bm{\mathcal{J}}$. In both the cases, formation of a phase-space vortex from two-stream instability has been simulated using Vlasov kinetic simulation employing the FS scheme. The hydrodynamic nature of the phase-space evolution shown in the figures portray the phase-space evolution as a fluid flowing along the positive time axis, deforming and evolving in accordance with the physics of interacting electric field and the plasma\cite{morse1969one}.

The vectorisation of the dynamical phase-space and accordingly the Vlasov equation presented in this section dispenses that the evolution of the phase-space df occurs independently along each direction of the dynamical phase-space. This orthogonal evolution of the phase-space implies that the evolution of the phase-space df (and the flux density $\bm{\mathcal{J}}$) can be simulated by individually evolving it along each direction, instead of shifting the evolution along two axis-sets, as has been done classically\cite{Cheng1976, Filbet2003}. Using this outcome, we devise and present a new numerical scheme for the dynamical evolution of the plasma phase-space in the next section.

\section{A Numerical Integration scheme for the Vlasov Equation and its stability analysis}\label{sec3}
As discussed above, the vectorised phase-space makes it convenient to numerically evolve the phase-space df along individual coordinates independently of each other. This is highlighted by the vector form of the Vlasov equation (\ref{vlasoveqninvectorform}). Expanding the differential to present the classical form of the VP system,
\begin{gather}\allowdisplaybreaks
    \bm{\nabla}\cdot \bm{\mathcal{J}} = \frac{\partial f}{\partial t} + v_x\frac{\partial f}{\partial x} - \frac{q}{m}\frac{\partial \phi}{\partial x}\frac{\partial f}{\partial v_x}=0.
    \intertext{In the above equation, the terms represent independent gradients along each axis of the dynamical phase-space. Discretizing the equation, we get --}
    \frac{\Delta f}{\Delta t}\Bigr|_{\Delta x=0, \Delta v_x = 0} + v_x\frac{\Delta f}{\Delta x}\Bigr|_{\Delta t =0, \Delta v_x =0} - \frac{q}{m}\frac{\partial \phi}{\partial x}\frac{\Delta f}{\Delta v_x}\Bigr|_{\Delta x=0, \Delta t =0}=0.\label{discreteform}
    \intertext{Evolving the dynamical phase-space coordinates in time $\Delta t$ as shown in equations (\ref{delta_t}), (\ref{delta_x}) and (\ref{delta_v}), we get --}
    \frac{\Delta f}{\Delta t}\Delta t + (v_x\Delta t)\frac{\Delta f}{\Delta x} + \left(-\frac{q}{m}\frac{\partial \phi}{\partial x}\Delta t\right)\frac{\Delta f}{\Delta v_x}= \Delta f\Bigr|_{t}^{t+\Delta t} +\frac{\Delta f}{\Delta x}\Delta x\Bigr|_{t}^{t+\Delta t} + \frac{\Delta f}{\Delta v_x}\Delta v_x\Bigr|_{t}^{t+\Delta t}=0 .\\
    \Rightarrow \Delta f\Bigr|_{t}^{t+\Delta t} + \Delta f\Bigr|_{x}^{x+\Delta x} + \Delta f\Bigr|_{v_x}^{v_x+\Delta v_x} =0.\label{algo_1}
\end{gather}
Equation (\ref{algo_1}) presents the complete evolution of the phase-space df in a time interval $\Delta t$ in terms of a set of flow-like progressions along each axis of the dynamical phase-space. This evolution, as shown in equation (\ref{algo_1}), occurs independently along each axis, as discussed above. It can also be seen that the equation represents a time-explicit integration scheme with one unknown. Using indices $n$, $j$ and $k$ for descretized steps of time $(c\Delta t)$, position $(\Delta x)$ and velocity $(\tau\Delta v_x)$, equation (\ref{algo_1}) obtains the form --
\begin{equation}\label{algo_2}
    f^{n+1}_{j,k} = f^n - \Delta_xf^n -\Delta_{v_x}f^n.
\end{equation}
Equation (\ref{algo_2}) presents a numerical algorithm to evolve the phase-space df in time. While the temporal evolution of the phase-space df itself is presented in terms of its flow along the position and velocity axes, the evolution of the position-velocity coordinates themselves can be determined directly using equations (\ref{delta_x}) and (\ref{delta_v}). The phase-space density at these points can be determined by numerical interpolation. The complete algorithm can be described in the following step, in correspondence to the FS scheme\cite{Cheng1976} shown in equation (\ref{cheng_knorr_scheme}):
\begin{gather}
    f(x,v_x,t+\Delta t) = f(x,v_x,t) + [f(x\rightarrow x-v_x\Delta t,v_x,t) - f(x,v_x,t)] + [f(x,v_x-qE(x)\Delta t/m, t) - f(x,v_x,t)],\\
    \Rightarrow f(x,v_x,t+\Delta t) = f(x\rightarrow x-v_x\Delta t,v_x,t) + f(x,v_x-qE(x)\Delta t/m, t) - f(x,v_x,t).\label{new_algro}
\end{gather}

In the above equations (\ref{discreteform} - \ref{new_algro}), $c$ and $\tau$ are normalised to be $1$. As can be seen, the numerical scheme described by equation (\ref{new_algro}) presents a one-step algorithm with two interpolation shifts, which can be performed simultaneously and are independent of each other. It can be seen that the numerical accuracy of the scheme depends on the accuracy of the chosen interpolation technique, and does not decrease beyond it since consecutive interpolations are not performed on the phase-space df. if $\Delta \epsilon$ is the numerical error produced in the phase-space df due to each interpolation, the FF scheme compounds this error by the the successive interpolations, as shown in equation (\ref{cheng_knorr_scheme}). 
\begin{equation}
    f^n\rightarrow f^* (= f + \Delta \epsilon) \rightarrow f^{**} (= f + \Delta\epsilon+\Delta\epsilon) \rightarrow f^{***} ( =f + \Delta \epsilon+ \Delta \epsilon+ \Delta \epsilon) = f^{n+1}.
\end{equation}
Therefore the numerical error at one time-step becomes $3\Delta\epsilon$. 
However, this compounding collective error does not occur in the ANI scheme, which does not utilise the successive interpolation algorithm.
\begin{equation}
    f^{n+1} = f^n_{j',k} + f^n_{j,k'} (=f + \Delta \epsilon) - f^n_{j,k}.
\end{equation}
This agile numerical integration (ANI) scheme is therefore an accuracy preserving, single-step semi-Lagrangian scheme, with comparatively more accuracy, even though the order of the accuracy remains same. The stability of the scheme can be analysed using the well-known Von-Neumann technique\cite{Crank1996AType, Charney1950NumericalEquation}. Representing the numerical error at a phase-space grid $(j,k)$ as $\epsilon_{j,k}^n$ for the $n^{\text{th}}$ time-step, and expressing it in terms of a Fourier series, we get 
\begin{gather}
    \epsilon(t,x,v_x) = \sum_{\sigma,\eta} A_{\sigma,\eta}(t) \exp(i k_\sigma x)\cdot \exp(ih_{\eta}v_x).\\
    \text{Here,}\quad 
    k_\sigma=\frac{\pi \sigma}{L} \quad \text{and} \quad \sigma = -\frac{L}{\Delta x}, -\frac{L}{\Delta x} +1,\quad .\quad . \quad 0 \quad .\quad .\quad \frac{L}{\Delta x} -1, \frac{L}{\Delta x}.\\
    \text{Similarly},\quad h_\eta = \frac{\pi\eta}{2 v_0} \quad \text{and} \quad \eta = -\frac{2v_0}{\Delta v_x}, -\frac{2v_0}{\Delta v_x} +1,\quad .\quad . \quad 0 \quad .\quad .\quad \frac{2v_0}{\Delta v_x} -1, \frac{2v_0}{\Delta v_x}.
    \intertext{Here, $L$ and $v_0$ represent the length and velocity amplitude of the system, and $A_{\sigma,\eta}$ is the amplitude of the error function. Therefore, we get --}
    \epsilon_{j,k}^{n+1} = A_{\sigma,\eta}(t+\Delta t) \exp(i k_\sigma x)\cdot \exp(ih_{\eta}v_x),\quad \text{where } t^{n+1} = t^n + \Delta t. \\
    \epsilon_{j',k}^{n} = A_{\sigma,\eta}(t) \exp(i k_\sigma (x-v_x\Delta t))\cdot \exp(ih_{\eta}v_x),\quad \text{where } x_{j'} = x_j - v_x\Delta t. \\
    \epsilon_{j,k'}^{n} = A_{\sigma,\eta}(t) \exp(i k_\sigma x)\cdot \exp(ih_{\eta}(v_x - q\vec{E}(x_j)\Delta t/m)),\quad \text{where } v_{x _{k'}} = v_{x_k} - \frac{q}{m}\vec{E}(x_j)\Delta t. 
\end{gather}
Inserting the above equations into the numerical scheme presented in equation (\ref{new_algro}), we get --
\begin{multline}\label{eqnoferrrors}
    A_{\sigma,\eta}(t+\Delta t) \exp(i k_\sigma x)\cdot \exp(ih_{\eta}v_x) = A_{\sigma,\eta}(t) \exp(i k_\sigma (x-v_x\Delta t))\cdot \exp(ih_{\eta}v_x) \\+ A_{\sigma,\eta}(t) \exp(i k_\sigma x)\cdot \exp(ih_{\eta}(v_x - q\vec{E}(x)\Delta t/m)) -  A_{\sigma,\eta}(t) \exp(i k_\sigma x)\cdot \exp(ih_{\eta}v_x).
\end{multline}
Equation (\ref{eqnoferrrors}) simplifies into --
\begin{equation}\label{stability_ratio}
     \frac{A_{\sigma,\eta}(t+\Delta t)}{ A_{\sigma,\eta}(t)} =  \exp({-ik_\sigma v_x\Delta t}) + \exp(-ih_\eta q\vec{E}(x)\Delta t/m) -1.
\end{equation}
The ratio derived in equation (\ref{eqnoferrrors}) describes the growth rate of numerical error at each time-integration and must be less than or equal to 1 for the numerical scheme to be stable. Expanding equation (\ref{eqnoferrrors}) be introducing $\theta_1 = k_\sigma v_x \Delta t$ and $\theta_2 = h_\eta q\vec{E}(x)\Delta t/m$,
\begin{gather}
    \exp(-i\theta_1) + \exp(-i\theta_2) -1 \leq 1\Rightarrow \exp(-i\theta_1) + \exp(-i\theta_2) \leq 2. 
    \intertext{Using the identity $e^{-i\theta} = \cos(\theta) - i\sin(\theta)$, we get --}
    \cos\theta_1 + \cos \theta_2 - i\left(\sin\theta_1 + \sin\theta_2\right) \leq 2 \Rightarrow \cos\left(  \frac{\theta_1 + \theta_2}{2} \right)\cos\left(  \frac{\theta_1 - \theta_2}{2} \right) -i\sin\left(\frac{\theta_1 +\theta_2}{2}\right)\cos\left(\frac{\theta_1 - \theta_2}{2}\right)\leq 1.
    \intertext{Taking $(\theta_1 +\theta_2)/{2} = \Theta_A$ and $(\theta_1 -\theta_2)/{2} = \Theta_B$ and finding the squared-modulus on both sides, we get --}
    \cos^2\Theta_A \cos^2\Theta_B + \sin^2\Theta_A \cos^2\Theta_B = \cos^2\Theta_B(\cos^2\Theta_A + \sin^2\Theta_A )\\
    =  \cos^2\Theta_B \leq 1,\quad \text{which is true for all values of } \Theta_B.\label{unconditional_stabiliity}
\end{gather}
Equation (\ref{unconditional_stabiliity}) therefore presents the unconditional stability of the numerical algorithm presented in this work in equation (\ref{new_algro}). In this section, a semi-Lagrangian, one-step ANI scheme for the time-integration of the Vlasov Equation has been introduced and its stability is discussed. As stated earlier, the numerical accuracy of this scheme is dependent on the interpolation technique. When coupled with the Poisson equation, the numerical accuracy then also depends on the choice of the Poisson solver. In the next section, we employ this ANI scheme in order to simulate some well-known phenomena in plasma physics, and compare the results obtained and the speed of the simulation with the FS scheme.

\section{Numerical Results and Discussions}\label{sec4}
\begin{figure}[!ht]
    \centering
    \includegraphics[width=0.7\linewidth]{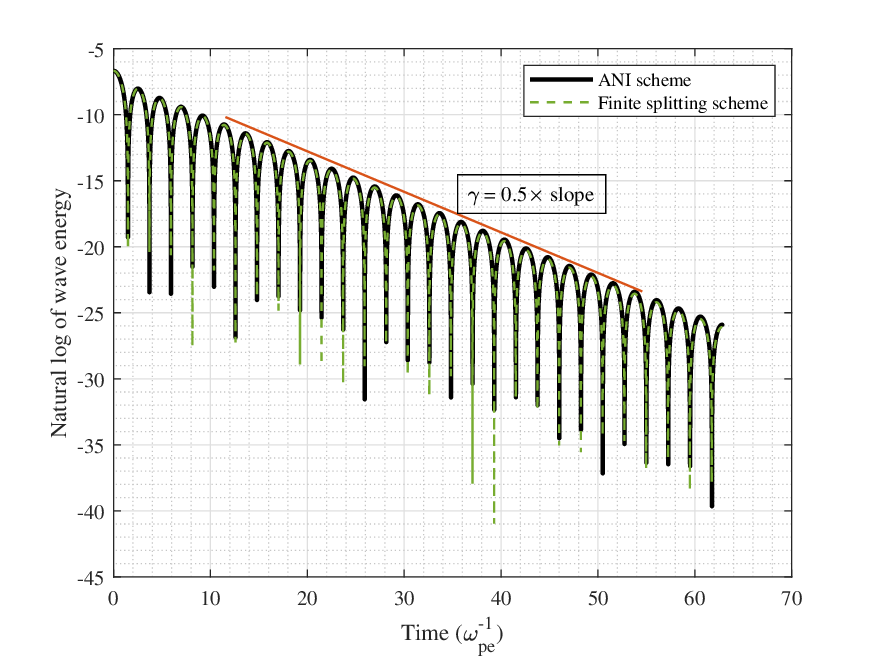}
    \caption{Linear Landau damping of electron Langmuir wave. (Wave energy vs. time plot. Graph representing damping simulation using the ANI scheme (black, continuous curve) and the finite-splitting scheme (green, dashed curve). Damping factor calculated by the slope of the red line.}
    \label{fig:Lldelectron}
\end{figure}

In this part of our work, we employ the one-step, ANI scheme presented in section \ref{sec3} equation (\ref{new_algro}) to numerically simulate some well-known collisionless plasma phenomena. This includes linear wave propagation and its Landau damping, and the plasma echo phenomena. These results are compared with the simulation results of the FS scheme\cite{Cheng1976}. We test the numerical accuracy and speed of our reduced scheme and present the results. In order to perform the numerical simulation, we employ the cubic spline interpolation in our scheme, which is well-known to exhibit higher-order accuracy up to the $4^{th}$ order. We employ a phase-space grid of size $M\times N$, where $M$ represents the number of grid points in the position $(x)$ space and $N$ represents the number of grid points in the velocity $(\tau v_x)$ space. We take $M=512$ and $N=512$, therefore dealing with a total of $2,62,144$ grid points. This ensures a small grid-size, leading to significantly reduced numerical dissipation. We specifically deal with the case of electron plasma waves, and simulate the electron phase-space. We normalise the position with electron Debye length $\lambda_{De} = \sqrt{\varepsilon_0 K_BT_e/n_0e^2}$, with $K_B$ and $T$ being the Boltzmann's constant and electron temperature, respectively. The electron charge is represented by $e$. Similarly, the particle velocity is normalised with electron thermal velocity $v_{Te}=\sqrt{2K_BT_e/m}$ and time with inverse plasma frequency $\omega_{pe} = \sqrt{n_0 e^2 / \varepsilon_0 m}$. Singly charged ions are assumed to remain stationary and form a uniform, neutralising background. The resultant VP system is as follows:
\begin{gather}
    \frac{\partial f}{\partial t} + v_x\frac{\partial f}{\partial x}-E(x)\frac{\partial f}{\partial v_x}=0,\label{normalised_Vlasov}\\
    \frac{\partial E}{\partial x} = - \frac{\partial ^2 \phi}{\partial x^2} = 1 - \int_{-\infty}^{\infty} fdv_x.\label{normalised_poissoneqn}
    \end{gather}
At each time-step, the Vlasov equation is numerically integrated and the Poisson equation is solved after integrating the modified phase-space df. We employ the inverse fast-Fourier technique in order to solve the Poisson equation, and use periodic boundary conditions. The following CFL conditions are employed for additional stability at each time-integration step\cite{Courant1928} --
\begin{gather}
    dt = \frac{\Delta x}{v_0} \quad\text{or}\quad dt = \frac{\Delta v_x}{E_0},\quad\text{whichever is smaller}.
\end{gather}
Here, $v_0$ is the maximum velocity and $E_0$ is the absolute electric field amplitude.
We start with the numerical simulation of linear Landau damping\cite{Landau1946OnPlasma} of an electron plasma wave, which is a collisionless decay in the wave energy occurring due to transfer of the wave energy to the resonant particles. For the same, the initial distribution of the electrons in phase-space is perturbed by adding an electron distribution as follows:
\begin{equation}\label{damping_perturbation}
    f(x,v_x,t=0) = \frac{1}{\sqrt{\pi}}\exp(-v_x^2)\cdot(1+A\cos(\kappa x)),
\end{equation}
    where $A$ is the perturbation amplitude and $\kappa$ is the wave number. We use a small amplitude of $A = 0.01$ in normalised units in order to restrict the damping in the linear domain. We choose wave number $\kappa$ as $0.5$. The values of observed wave angular frequency $\omega$ and damping factor $\gamma$, along-with the time of the simulation in both finite-splitting scheme case and the ANI scheme are reported in table \ref{table:linear_electron_damping0.5}. The observation has been shown in fig. \ref{fig:Lldelectron}. For the simulation, the same computation set-up has been used for both numerical schemes and the time of the complete code-run has been calculated digitally, accurate to the third decimal. Three runs of the codes have been used and the average of the simulation times of the three trials has been reported for each case.

  \begin{table}[ht]
\caption{Data of numerical simulation of linear Landau damping of electron plasma wave. Comparison of numerical simulation schemes in terms of accuracy and speed for wave angular frequency $\omega$ and damping factor $\gamma$ are reported for $\kappa = 0.5$.} 
\label{table:linear_electron_damping0.5}
\begin{center}       
\begin{tabular}{|l|l|l|l|l|} 
\hline
\rule[-1ex]{0pt}{3.5ex}  & $\kappa$ & $\omega $ (error) & $\gamma$ (error)& simulation time (seconds) \\
\hline
\rule[-1ex]{0pt}{3.5ex} Theoretical&0.5 & 1.4156 & -0.153359 & -- \\
\hline
\rule[-1ex]{0pt}{3.5ex} FS scheme &0.5 &1.4159 (0.02119\%) & -0.152356 (-0.653\%)& 862.601 \\
\hline
\rule[-1ex]{0pt}{3.5ex} ANI scheme &0.5 &1.4158 (0.01412\%)& -0.153264 (-0.061\%)& 469.441 \\
\hline
\end{tabular}
\end{center}
\end{table}

\begin{figure}[!ht]
    \centering
    \includegraphics[width=0.7\linewidth]{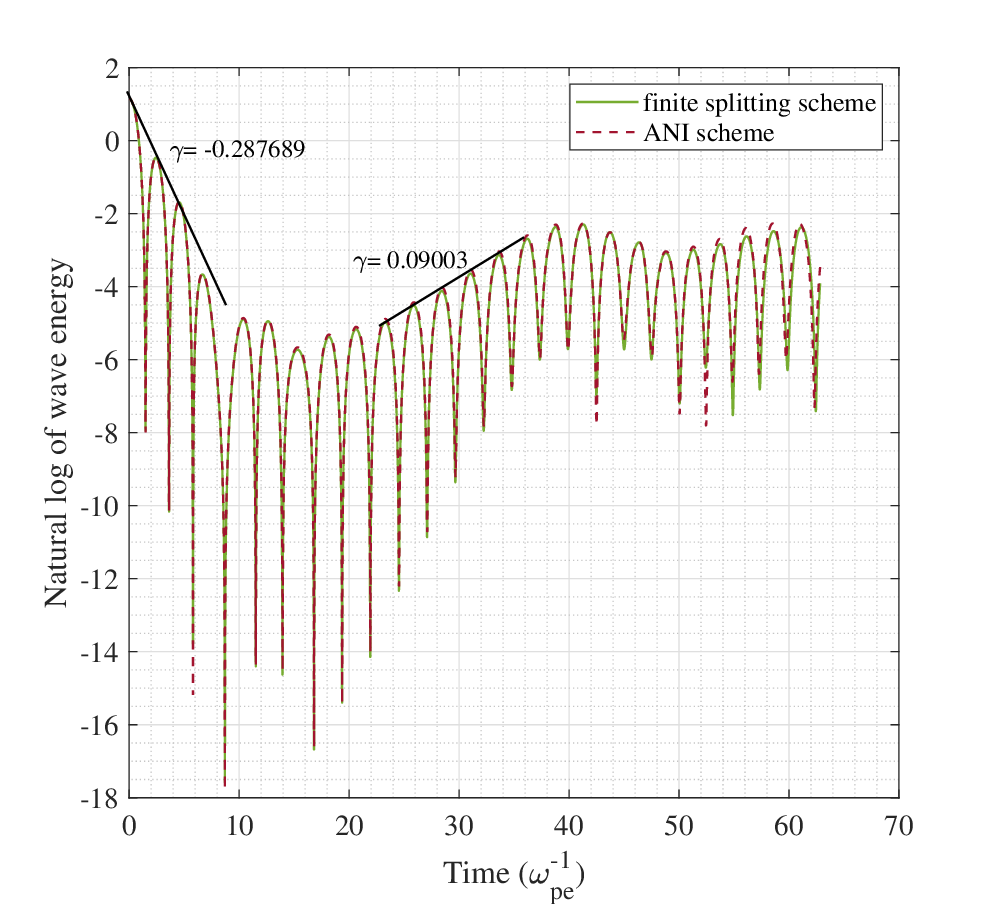}
    \caption{Nonlinear Landau damping of Langmuir waves, with amplitude $A=0.5$ and wave-number $\kappa=0.5$. Wave damping rate and growth rate shown. Simulation performed using both ANI (red, dashed curve) and FS (green, continuous curve) schemes.}
    \label{fig:nonlinear0.5}
\end{figure}

\begin{figure}
    \centering
    \includegraphics[width=0.7\linewidth]{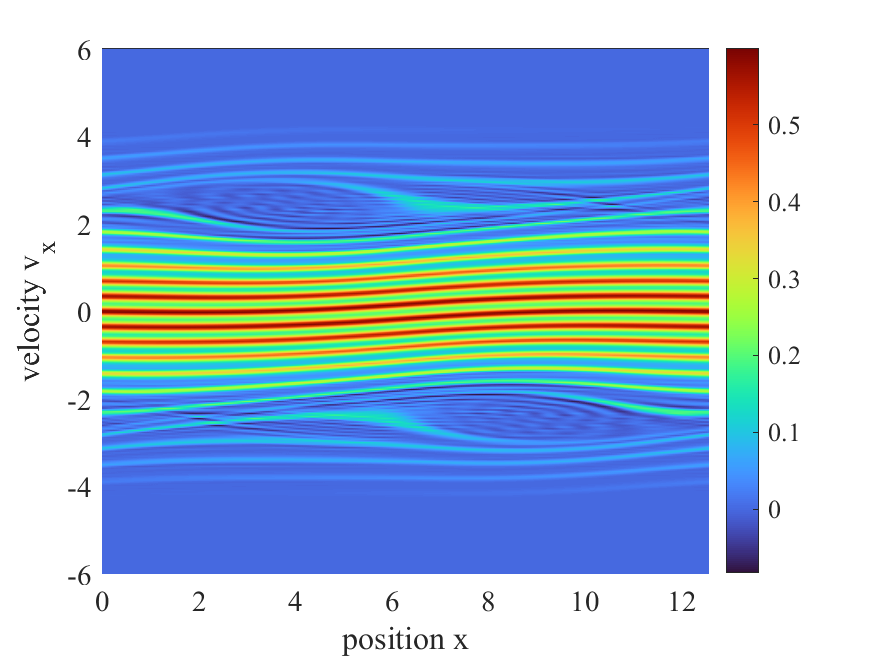}
    \caption{Phase-space density portrait exhibiting presence of electron phase-space holes due to nonlinear Landau damping. Simulation performed using the ANI scheme with $A=\kappa=0.5$. Colour gradient representing magnitude of the phase-space df.}
    \label{fig:holesoflandau}
\end{figure}

From the results presented in table \ref{table:linear_electron_damping0.5}, it is clear that the ANI scheme developed herein reduces the computational time significantly, to almost half when compared with the FS scheme. The accuracy of the two schemes are comparable to each other. For the nonlinear Landau damping case, the wave perturbation amplitude $A$ is increased to $0.5$ to shift the interaction into nonlinearity. This is the transfer of energy initially from the wave to the slower resonant particles, which then accelerate beyond the wave phase-speed. This is followed by the transfer of energy from these higher speed particles back to the wave, thus presenting an oscillation of the wave energy and formation of phase-space df troughs centered at the wave phase-speed. Using $\kappa = 0.5$, we simulate the electron phase-space. Table \ref{table:nonlinear_electron_damping0.5} and fig. \ref{fig:nonlinear0.5} presents the results. The wave energy can be seen to decay initially with a decay factor $\gamma = -0.575377$ and then grow with a growth factor $\gamma = 0.180057$, to slowly reach a saturation. This is a typical behaviour of nonlinear collisionless damping and has been reported in many previous works \cite{Cheng1976, Manfredi1997}. These results are in agreement with the data reported by Gazdag \cite{Gazdag1975NumericalMethod} and later by Cheng and Knorr\cite{Cheng1976}, with errors of $\leq 0.236\%$. Fig. \ref{fig:holesoflandau} shows the formation of phase-space holes due to the nonlinear trapping of electrons in the nonlinear Landau damping phenomena. These are regions of reduced particle densities in the phase-space, travelling in either directions with the wave phase-speeds.

\begin{table}[ht]
\caption{Data of numerical simulation of nonlinear Landau damping of electron plasma wave, reported for $\kappa = 0.5$ and $A = 0.5$. Decay and growth rates of the electrostatic wave amplitude reported, along-with the simulation times of the finite-splitting and ANI scheme.} 
\label{table:nonlinear_electron_damping0.5}
\begin{center}       
\begin{tabular}{|l|l|l|l|l|} 
\hline
\rule[-1ex]{0pt}{3.5ex}  $\kappa$ &  $\gamma$ (decay) & $\gamma$ (growth) & simulation time FS (seconds)  & simulation time ANI (seconds)  \\
\hline
\rule[-1ex]{0pt}{3.5ex} 0.5 & -0.287689& 0.0900289 & 783.298 & 404.373\\
\hline
\end{tabular}
\end{center}
\end{table}

We next present the numerical simulation of the temporal plasma echo phenomena\cite{Gould1967PlasmaEcho, Malmberg1968PlasmaExperiment} using the ANI scheme. Plasma echo is well-known to be a highly nonlinear phenomena which is quite difficult to simulate numerically due to its dependence on the numerical stability\cite{Hou2011}. This wave-wave interaction phenomena occurs due to the persistence of the electron velocity df oscillations even after the Landau damping of the waves' macroscopic properties. Thus, two completely damped waves exhibit resonant interactions resulting in higher order oscillations after some measurable time intervals. 

This time interval depends on the wave-nature of the two waves. For initial perturbations of the form represented in equation (\ref{damping_perturbation}), presented in the plasma after a time interval of $\mathcal{T}$, having wave-numbers $\kappa_1$ and $\kappa_2$, harmonic modes $\alpha_1$ and $\alpha_2$ and amplitudes $A_1$ and $A_2$ respectively, the echo occurs at a time interval $ t_{echo}$ after the first perturbation, such that
\begin{equation}\label{echo_time}
     t_{echo} = \frac{\alpha_2\kappa_2}{\alpha_2\kappa_2 - \alpha_1\kappa_1}\mathcal{T}.
\end{equation}
For the simulation of the plasma temporal echo, the velocity grid spacing of the phase-space should be small enough for recurrence effect to occur much later than the echo time. The recurrence time $T_R$ is equal to $2\pi/k\Delta v_x$. The plasma echo phenomena is also demonstrated to be dependent on the nature of the damping of the waves\cite{Hou2011}. In case of nonlinear damping, due to particle trapping, the initial oscillations produced by the perturbations undergo phase-mixing, resulting in loss of initial information of the waves. This causes plasma echoes to disappear as the nonlinearity increases in the damping nature. Hence, small amplitude perturbations must be used.

\begin{figure}
    \centering
    \includegraphics[width=0.7\linewidth]{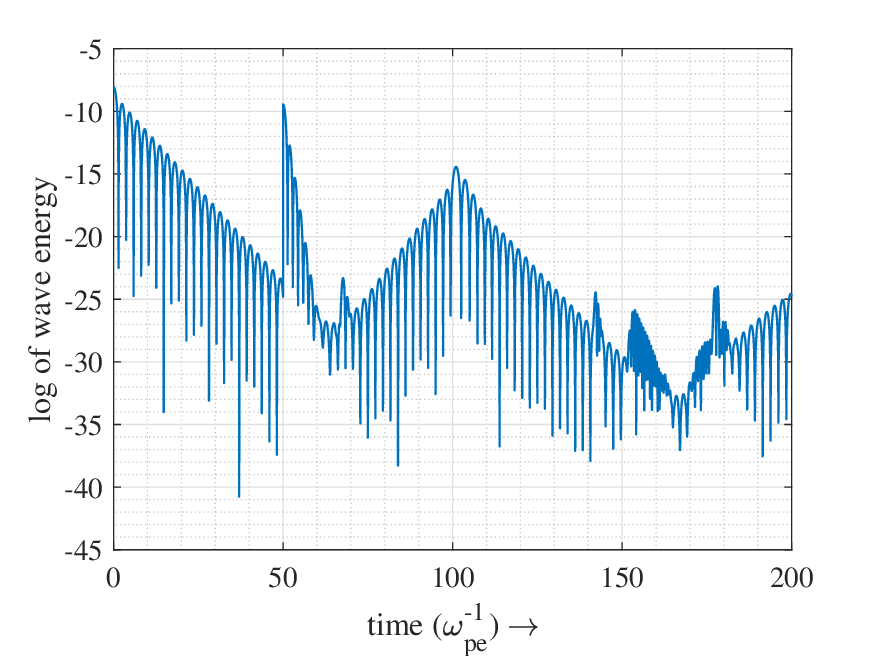}
    \caption{A temporal plasma echo. Wave amplitudes $A_1=A_2 = 0.005$ and $\kappa_2 = 2\kappa_1 = 1.0$ (both in $1^{st}$ harmonics). Echo occurring at time $t_{echo}=100.918\omega_{pe}^{-1}$, as against the theoretical prediction of $t_{echo}=100.918\omega_{pe}^{-1}$ (error $= 0.918\%$. }
    \label{fig:echo_hua}
\end{figure}

For the simulation of temporal plasma echo, we introduce two perturbations similar to the form represented by equation (\ref{damping_perturbation}). We use a small amplitude of $A_1=A_2 = 0.005$ in order to restrict the damping to the linear case. We use the waves of wave numbers $\kappa_1 = 0.5$ and $\kappa_2=1.0$ and introduce them at a time interval $\mathcal{T}$ of $50.0\omega_{pe}^{-1}$, at their natural (first) harmonic cases. From equation (\ref{echo_time}), the predicted echo occurrence time $t_{echo}=100.0\omega_{pe}^{-1}$. Fig. \ref{fig:echo_hua} presents the numerical simulation of the temporal plasma echo using the ANI scheme. It can be seen that in the numerical simulation, the plasma temporal echo occurs at time $t_{echo}=100.918\omega_{pe}^{-1}$ (represented by the peak of the wave energy).
The accuracy of the ANI scheme is evident with an error in the echo time of $0.918\%$, as compared to the FS scheme which presents an error of $1.732\%$. The simulation time using ANI scheme is found to be $2023.415$ sec., as compared to the FS scheme ($3940.127$ sec). The later peaks formed in the simulation, as visible fig. \ref{fig:echo_hua} occur due to the numerical recurrence phenomena\cite{Hou2011}. 

In this section, the accuracy, speed and applicability of the ANI scheme has been established in order to numerically simulate various linear and nonlinear plasma phenomena using the kinetic Vlasov simulation. In the next section, we conclude our work by briefly discussing the work presented in this article.

\section{Conclusion}\label{sec5}
In this article, the phase-space of collisionless electrostatic one-dimensional plasmas has been presented and discussed as a three-dimensional vector space by observing its evolution in time. The phase-space evolution is then showcased as a hydrodynamic flow described by the phase-space velocity field, defined in equation (\ref{u}) and the flux density vector field, defined in equation (\ref{j_Value}). This flow-like evolution presents an intuitive, though unconventional outlook of the phase-space dynamics of the system. Using this approach, the formation of coherent kinetic structures and phase-space contortions can be analysed as fluid-analogous behaviour. 

The vectorisation of the dynamic phase-space also permits the development of a numerical scheme for the Vlasov equation integration. This new agile numerical integration (ANI) scheme, as developed in this work, is an accurate and stable numerical scheme for the Vlasov equation and reduces the computation burden to almost half, when compared to the finite splitting scheme\cite{Cheng1976}, which is a benchmark numerical technique for the kinetic Vlasov simulation. It has been shown in this work that this new ANI scheme reduces the computational times to almost half, when compared with the finite splitting scheme. It has high accuracy which is comparable to the finite splitting scheme, and is capable to simulate both linear and nonlinear phenomena occurring in the plasma.

The ANI scheme can be employed in various numerical simulations of the plasma kinetic theory to study collisionless plasma phenomena. Due to its reduced computation load, it can be utilised for faster simulations in less-powered computation set-ups. In this work, we have employed the cubic spline interpolation technique in the numerical simulations, which has a high accuracy. However, other interpolation techniques can be employed to further enhance the speed or accuracy of the numerical simulation. The scheme exhibits unconditional stability, unlike the conservative Eulerian scheme which depends on the spatio-temporal grid sizes. Hence, the numerical scheme presents a fast and accurate numerical integration technique of the Vlasov integration without compromising on the stability of the simulation.


\bibliography{references} 
\bibliographystyle{spiebib} 

\end{document}